\documentstyle[12pt]{article}

\begin{document}
\begin{titlepage}
\today          \hfill
\begin{center}
\hfill    OITS-698 \\

{\large \bf
A Note 
on Gaugino Masses in
Kaluza-Klein/Radion Mediated SUSY Breaking
}
\vskip .15in
Kaustubh Agashe

{\em
Institute of Theoretical Science \\
University
of Oregon \\
Eugene OR 97403-5203}

\vskip .15in

email: agashe@neutrino.uoregon.edu

\end{center}

\vskip .05in

\begin{abstract}

We review the equivalence of 
two approaches to study 
theories with gauge fields in extra spatial dimensions,
namely the ``$4D$'' approach (with KK states) and the ``$5D$''
approach (with matching to the $4D$ theory at the
compactification scale).
In particular, we reiterate that there are
two different power-law scalings of ``effective'' gauge couplings. 
In a supersymmetric framework with SUSY breaking
in the radius modulus, i.e., the field
which fixes the size of the extra dimensions, these
two approaches seem to give
gaugino masses at loop-level
(with a possible
enhancement due to large number of Kaluza-Klein states) \cite{ky}, and 
tree-level
\cite{cl}, respectively.
We show explicitly
how
this discrepancy can be resolved.

\end{abstract}

\end{titlepage}

\newpage
\renewcommand{\thepage}{\arabic{page}}
\setcounter{page}{1}

\section{Introduction}

There are two approaches to analysing theories with SM gauge fields in
extra spatial dimensions \cite{anto}: \\
1)
``$4D$'' approach
in which the extra dimensions ``appear'' in the form of Kaluza-Klein (KK)
excitations of gauge fields.
In this approach, the
``effective'' gauge coupling 
at energy scale $E$ $( > R^{-1})$
is $N_{KK} (E) 
\times g_{4D}^2 (E)$, where $N_{KK} (E) \sim R^{\delta} E^{\delta}$
is number of
KK states lighter than $E$ (including the zero-modes).
Here $\delta$ is the number of extra dimensions
and $R$
is a typical size of an extra dimension. 
For a non-abelian gauge group,
this effective coupling 
(i.e., number of KK states in loop growing
with energy) results in
$g_{4D}$ running with power of energy
\cite{ddg}. \\ 
and \\
2)
``$(4 + \delta) D$'' or for short
``$5D$'' approach in which the gauge fields are treated as effectively
being 
in $(4 + \delta) D$ (non-compact) dimensions
(above the compactification scale $\sim R^{-1}$) 
followed by matching to the (effective)
$4D$ theory (at $R^{-1}$) given by
$g_{4D}^2 \sim g_{( 4 + \delta ) D}^2 / R^{\delta}$.
In this approach, the
$( 4 + \delta ) D$ gauge coupling should ``run''
with power of energy since the gauge coupling
is dimensionful 
and there is an integration over virtual
extra-$D$ momentum which results in a power-divergence.

Of course, these two approaches should be 
equivalent. However, 
in one example, this equivalence is
not clear. 
%(at least to the present author). 
Consider a supersymmetric version of this framework
in which the radion, i.e.,
the scalar field which
determines the size $R$ of the extra dimensions and  
hence the mass $\sim n / R$ of the KK states,
is part of a chiral superfield.
Suppose $F$-component of the radion (chiral superfield)
has a vev and thus breaks SUSY. 
In this case, as discussed recently,
SUSY breaking is mediated
to the MSSM gauginos (which propagate in the bulk) by 
the KK states
\footnote{KK states 
have a non-supersymmetric mass spectrum
since their masses $\sim n / R$ are
determined by the radion.} \cite{ky} (KK mediated SUSY breaking: KKMSB)
or (what should be equivalent) by coupling of gauge
fields to the radion
\cite{cl} (radion mediated SUSY breaking:
RMSB).  
This contribution to gaugino mass is determined by
how the gauge 
coupling at low energies depends on $R$ (i.e.,
the radion) -- a priori, it is not clear that 
this dependence is the same in the two approaches. 
In fact, 
the $4D$ approach {\em as used in} \cite{ky}
gives gaugino masses 
at loop-level, 
%(possibly enhanced by
%large number of KK states) 
whereas \cite{cl} uses the $5D$ approach
to obtain gaugino masses at tree-level -- these two results
are obviously different.

In this paper, we review the
equivalence of 
the two approaches, in particular, the notions of
``quantum'' and ``classical'' power-law scalings of effective
gauge couplings.
This discussion is then used in the last section 
to clarify and resolve the discrepancy
between the above two results for gaugino masses in KKMSB/RMSB. We 
show 
that, even in the $4D$ approach, there {\em is} a tree-level contribution to
gaugino masses (corresponding
to the tree-level effect in 
the $5D$ approach of  
\cite{cl}). This is due to the dependence of 
$g_{4D}$ {\em at the cut-off} on $R$.
However, this effect was
%{\em assumed to be absent} 
{\em neglected} in \cite{ky}. Also, when 
the
$( 4 + \delta ) D$ theory is strongly coupled
or (equivalently) when there are a large number of KK states (in the $4D$
approach), the loop
contributions to gauge coupling 
(and to gaugino mass) studied in \cite{ky} 
(and {\em not} considered in the {\em tree-level} analysis of \cite{cl}) 
can be, {\em a priori}, {\em comparable} to tree-level terms.
However, it turns out (as we show) that, even in this case, the net 
result 
for gaugino masses
(i.e., combining
tree-
and loop-level effects) is well-approximated
by the (tree-level) expression
given in \cite{cl}, provided $g_{4D} \stackrel{<}{\sim} O(1)$
(as in the MSSM case).

\section{Power-Law Scalings of Gauge
Couplings}
We begin with a discussion of 
two different kinds of power-law scalings for effective
gauge couplings.
Both scalings have the same origin -- they are due to
number of
KK states increasing with energy
(in $4D$ approach) or dimensionful gauge coupling
and integration over
extra-$D$ momentum
(in $5D$ approach). As we explain below,
the terminology ``quantum'' or
``classical'' is used for this effect depending on
whether KK states (or integration over
extra-$D$ momentum in $5D$ approach) renormalize (at loop-level) the
gauge coupling or not.

\subsection{
``Classical'' Power-Law Scaling of Gauge Coupling}
\label{classical}
Consider $U(1)$ gauge field 
(``photon'') in bulk with all matter fields (``electron'') charged under 
$U(1)$ in $4D$ (i.e., on a $3$-brane).
Let us look at the
effect in
a tree-level 
process in the $4D$ approach, say, $e^+ e^- \rightarrow
\sum _{n, \; n^{\prime}} \gamma ^{(n)} \gamma
^{(n^{\prime})}$, where $\gamma ^{(n)}$ denotes the KK state
of the photon
with momentum in the extra dimension (and hence mass) $\sim n/R$.
For each $n$, $n^{\prime}$,
$\sigma \left( e^+ e^- \rightarrow
\gamma^{(n)} \gamma
^{(n^{\prime})} \right) \sim e_{4D}^4
/(16 \pi) \; 1 / E^2$ 
(as usual, where $E$ is the c.m. energy)
but the number of (kinematically accessible)
final-states
$\sim 
N^2_{KK} (E)$, where $N_{KK} (E) \sim
E^{\delta} R^{\delta}$ is the number of KK states
for each final-state photon. Thus, the total cross-section 
$\sim \left[ e_{4D}^2 N_{KK} (E) \right] ^2 
/(16 \pi) \; 1 / E^2$ can
grow with power of energy for large $\delta$.

This suggests that we can define an  
``effective'' gauge coupling (as mentioned in the introduction)
at energy $E (> R^{-1})$ which 
grows with power of energy:
\begin{equation}
e_{4D}^{\hbox{eff} \; 2} (E) 
\equiv e_{4D}^2 \times N_{KK} (E).
\end{equation}

Next, consider a loop-level process such as wavefunction
renormalization of electron -- at one-loop this also scales 
like a power of energy as
follows.
In the electron self-energy diagram, each photon KK state gives 
a log-divergent contribution  
(as usual from the $4D$ loop-momentum integration), but
we have to sum over
an infinite number of KK states 
so that
the $4D$ theory
with KK states appears non-renormalizable.
We can introduce a cut-off $\Lambda _{KK}$ to truncate the KK tower
and get a renormalizable (i.e., with {\em finite}
number of KK states) theory
as an approximation -- in this approximation, we can continue
to use the $4D$ language of energy-{\em dependent} or running
coupling/wavefunction. 
To compute the running 
electron wavefunction $Z_e$,
we introduce
the KK states as thresholds, i.e., 
we neglect the effect of KK states
heavier than the renormalization group (RG) scale (as is usually
done for other particle states).
So, in this
approximation, the effective coupling in the
renormalization group equation 
(RGE) for $Z_e$ 
at one-loop is (as above)
\begin{equation}
\frac{ e_{4D}^{\hbox{eff} \; 2} (E) }{ 16 \pi^2 } 
= \frac{ e_{4D}^2 }{ 16 \pi^2 } \times N_{KK} (E)
\label{eeff}
\end{equation}
instead of just $e_{4D}^2 / \left( 16 \pi^2 \right)$
(here $E$ is the RG scale).
Because of this effective coupling, 
the one-loop wavefunction renormalization
of electron runs with power of energy.

Although there is power-law
running of $Z_e$, running of $e_{4D}$ is
still logarithmic at one-loop (as in the case with
photon in $4D$) -- for simplicity, we will neglect this logarithmic 
(``mild'') dependence
of $e_{4D}$ on energy
in some cases (as in Eq. (\ref{eeff}))
and assume that $e_{4D}$ is constant.
This is because
matter fields (electron) do not
have KK states and so the coupling in vacuum polarization diagram
for photon
is still $e ^2 _{4D}$ and {\em not} $e ^2 _{4D} 
\times N_{KK} (E)$.
%\footnote{
In other words, $e^{\hbox{eff}}_{4D}$ does
{\em not} contribute to 
the photon wavefunction, $\Pi ^{\gamma \gamma} (q)$, 
and hence to $e_{4D}$, unlike the case of $Z_e$. Here
$\Pi _{\mu \nu} (q) \sim ( q_{\mu} q_{\nu}
- g_{\mu \nu} q^2 ) \times \Pi (q)$, where $q$ is the external ($4D$)
momentum and $\Pi _{\mu \nu} (q)$ is the gauge boson self-energy.
%}
Because $e_{4D}
^{\hbox{eff}}$, and hence processes involving the gauge coupling, 
acquire a $E^{\delta}$ dependence at {\em tree-level}
(for example, $e^+ e^- \rightarrow \gamma \gamma$ 
discussed above),
whereas
running of $e_{4D}$ is {\em not} affected
we call this 
a ``classical'' power-law scaling of
gauge coupling, although it does effect
$Z_e$ at loop-level.

From $( 4 + \delta ) D$ 
point of view, $e_{( 4 + \delta ) D}^2$ 
(at the vertices in the electron self-energy
diagram) has dimension of (mass)$^{- \delta}$
so that to obtain (dimension{\em less}) 
energy-{\em dependent}, i.e.,
running,
wavefunction 
renormalization of the electron,
we need to multiply by 
$E^{\delta}$ 
%\footnote{
-- in other words, by dimensional analysis,
the RGE for $Z_e$
is $ d \left( \ln Z_e (E) \right) / d ( \ln E ) \sim 1 / \left( 16
\pi^2 \right) e_{( 4 + \delta ) D}^2 E^{\delta}$.
%}
Explicitly, there is an
{\em extra} (compared to $4D$)
{\em loop}-momentum integration for the photon 
(corresponding to
extra-$D$ momentum) which gives power-divergence. This
implies that the $( 4 + \delta ) D$ theory is non-renormalizable.
%\footnote{
Thus,
there is also a contribution $\propto e_{( 4 + \delta ) D}^2 \; \Lambda
_{4 + \delta} ^{\delta}$,
where $\Lambda _{4 + \delta}$ is the cut-off of the $(4 +
\delta) \; D$ theory.
It is clear that 
this power divergence corresponds
to infinite sum over KK states in $4D$ approach,
i.e., $\Lambda_{KK} \leftrightarrow \Lambda _{4 + \delta}$.
%}
On the other hand, at one-loop,  
$e_{( 4 + \delta ) D}$ does not ``run'' 
like a power of energy since there is {\em no}
extra-$D$ momentum integration (for virtual
electrons) in the vaccum polarization
diagram for the photon. 
%\footnote{
In this diagram,
$e_{( 4 + \delta ) D}$ at vertices 
is also dimensionful, but the dimension is ``soaked'' up 
by factors of $R$ 
coming from the probability that the gauge boson
propagating in the extra dimensions of size $R$
is ``near'' the $3$-brane (where the interaction with electrons
takes place).
%} 
Of course, the $4D$ loop-momentum integration for electrons
results in the usual log-divergence.

In the process 
$e^+ e^- \rightarrow \gamma \gamma$, in the $5D$ approach, we 
again have dimensionful 
couplings 
and so to get correct dimension
for cross-section, we have to
multiply by powers of energy $\sim \left( E ^{\delta} \right) ^2$
(in addition to those in a $4D$ calculation)
-- this corresponds to 
integration over {\em real} extra-$D$ momentum 
(phase space) of each final-state photon.

\subsection{ ``Quantum'' Power-Law Scaling of Gauge Coupling}
This is absent for $U(1)$ case.
For non-abelian case
(again with ``quarks'' in $4D$), in the $4D$ renormalizable approach
mentioned above, 
the ``gluon'' wavefunction at one-loop 
%($\Pi ^{gg} (q)$,
%after subtracting the
%log-divergence) 
has power-law energy dependence
(unlike the photon wavefunction)
due to number of 
KK gluons in the loop growing with energy
(just like the electron wavefunction mentioned above),
i.e., the coupling for the vacuum polarization diagram
involving gluons in the loop {\em is} (effectively)
$g^2_{4D} \times N_{KK} (E)$
instead of $g_{4D}^2$ as in the photon case.
%\footnote{
In other words,
$g^{\hbox{eff}}_{4D}$ {\em does} contribute to
$\Pi ^{gg} (q)$ and hence renormalizes
$g_{4D}$, unlike in the $U(1)$ case.
%}
This implies that at one-loop
$g_{4D}^2 (E)$ 
``runs'' with a power of energy \cite{ddg}.
\footnote{In the $4D$ 
theory with {\em finite} number of KK states (and with 
decoupling of KK states heavier than the RG scale),  
we can call the power-law energy dependence
(at one-loop) of $g_{4D}$ as power-law ``running''
(we already used this language for the electron wavefunction above),
even though
the fundamental $(4 +
\delta) D$ theory is non-renormalizable.}
We will show this RG calculation in the renormalizable
approximation in section \ref{cut-off}.
Because $g_{4D}$, and hence $g_{4D}^{\hbox{eff}}$, 
depends on a power of energy {\em at one-loop}, we refer to this effect
as ``quantum'' power-law scaling of gauge coupling.
%\footnote{
To repeat, both classical and 
quantum power-law scalings of gauge coupling have the same origin
-- the number of KK states growing with a power of energy. The 
difference
is that the former refers to
power-law scaling of $g_{4D}^{\hbox{eff}}$ at
{\em tree-level}, whereas the latter refers to 
power-law scaling of $g_{4D}$ or $g_{4D}^{\hbox{eff}}$ at {\em (one)-loop
level}.
%}
Also, the effective coupling in
the RGE for quark wavefunction renormalization $Z_q$
is $N_{KK} (E) \times g_{4D}^2 (E) / \left( 16 \pi^2 \right)$,
where {\em both} $N_{KK}$ and $g_{4D}$ have power-law dependence on
$E$.

In $5D$ approach also, 
(dimensionless) wavefunction renormalization of gluon 
from the vacuum polarization
diagram \footnote{i.e.,
$\Pi ^{gg} (Q)$,
where $Q$ is the external ($( 4 + \delta ) D$)
momentum}, and hence $g_{( 4 + \delta ) D}$ (at one-loop),
depends on a 
power of energy (unlike photon case). The reason is that
$g_{( 4 + \delta ) D}^2$ at the vertices is dimensionful and there {\em is} 
extra-$D$ momentum integration for gluons (but not for quarks)
in the loop which changes the usual ($4D$)
log-divergence into a power-divergence (as in the case of $Z_e$). 
\footnote{$\Pi ^{gg} (Q)$ also
has a contribution $\propto g^2_{( 4 + \delta ) D} \; \Lambda 
_{4 + \delta} ^{\delta}$.} By dimensional analysis, the
RGE for quark wavefunction is $d \left( \ln Z_q (E) \right) 
/ d \left( \ln E \right)
\sim g_{ ( 4 + \delta ) D}^2 (E) \; E^{\delta}$ (up to 
a dimensionless loop-factor).

The above examples
just illustrate the well-known facts that 
a $( 4 + \delta ) D$ theory {\em with a cut-off
scale} is equivalent to a
$4D$ theory with a {\em finite} number of KK states
below this scale and that the sum over
KK states
corresponds to integration over extra-$D$ momentum.

\section{Matching the $4D$ and $( 4 + \delta ) D$ Theories}
On the basis of these arguments, we get the following plausible
translation dictionary
between the gauge couplings of the
$( 4 + \delta ) D$ theory and the 
$4D$ theory (with KK states), 
including {\em both} power-law
scalings:
\begin{eqnarray}
g^{\hbox{eff} 
\; 2}_{4D} (E) & \equiv g^2_{4D} (E) N_{KK} (E) & \sim
g_{( 4 + \delta ) D}^2 (E) \; E^{\delta}.
\label{dict}
\end{eqnarray}
Of course, this is valid for $E > R^{-1}$.
To repeat,
$g_{( 4 + \delta ) D}^2$ is dimensionful and hence by
dimensional analysis we
have to multiply by power of energy on the $( 4 + \delta ) D$ side
for ``comparing'' it to the $4D$
side; 
this power of energy
corresponds to extra-$D$ momentum integration
(either in the loop while evaluating 
wavefunction
renormalization or in an external leg as in
the process $e^+ e^- \rightarrow \gamma \gamma$). 
On the $4D$ side,
$N_{KK}$ (and hence $g^{\hbox{eff}
\; 2}_{4D} (E)$) grows with a power of
energy which matches $E^{\delta}$ on the $( 4 + \delta ) D$ side:
we refer to this as
classical power-law scaling.
Thus, this relation is easily justified 
at tree-level (i.e., without
$E$ dependence in $g_{4D}$ and
$g_{ ( 4 + \delta ) D}$) for both $U(1)$ and non-abelian gauge groups (see
the 
discussion of 
$\sigma \left( e^+ e^- \rightarrow
\gamma \gamma \right)$ and RGE for $Z_e (E)$ 
above).
For the $U(1)$ case, due to matter fields on a $3$-brane, 
there is the usual logarithmic running of $e_{4D}$ and we also
expect logarithmic dependence on energy
in $e_{( 4 + \delta ) D}$.
Thus, in the $U(1)$ case,
the above relation is fairly accurate
at the quantum-level also (i.e., {\em including} the 
logarithmic (mild) energy 
dependences in $e_{4D}$ and
$e_{( 4 + \delta ) D}$).

Furthermore, in the {\em non-abelian} case,
at the quantum-level,
$g_{4D}$ runs like a power of energy 
(quantum power-law scaling) as mentioned
earlier and as will be shown explicitly by a 
calculation in the renormalizable approximation (see section \ref{cut-off}).
As argued earlier,
$g_{( 4 + \delta ) D}$ also has (loop-suppressed) power-law dependence on
$E$ and
this dependence should correspond to the 
power-law running of $g_{4D}$. Thus the above relation is
plausible in the non-abelian case at the quantum level also, i.e.,
{\em including} the (power-law) energy dependences in $g_{4D}$ and 
$g_{( 4 + \delta ) D}$ (see the discussion of RGE for $Z_q (E)$ above). 
We can then say that $g_{( 4 + \delta ) D}$ also
``runs'' (at one-loop)
like a power of energy,
even though,
as mentioned earlier, the $( 4 + \delta ) D$ theory is
non-renormalizable and
so $g_{( 4 + \delta ) D}$ does not run in the $4D$ sense.
%\footnote{
In other words,
by dimensional analysis,
the {\em one-loop} ``RGE'' for $g_{( 4 + \delta ) D}$ is
$d g^{-2}_{( 4 + \delta ) D } (E) / d ( \ln E )
\sim 
E^{\delta}$ (up to a
dimensionless loop-factor).
%}
%\footnote{
The
$5D$ approach, i.e., the analysis with 
the dimension{\em ful}
coupling $g_{(4 + \delta) D}$, is similar to
the discussion in \cite{polchinski} of Wilsonian RGE's in $4D$ with  
{\em non}-renormalizable (irrelevant)
operators.
%} 
To prove
this correspondence between (one-loop)
running of $g_{4D}$ and that of
$g_{( 4 + \delta ) D}$,
and hence the above relation
in the non-abelian case, we would have to compute 
explicitly the loop correction in 
$( 4 + \delta ) D$ which will not be attempted here.

From Eq. (\ref{dict}) and using
$N_{KK} (E) \sim R^{\delta} E^{\delta}$,
we get the matching condition valid at {\em all} energies 
above $R^{-1}$:
\begin{equation}
g_{4D}^2 (E) \sim \frac{ g_{( 4 + \delta ) D}^2 (E) }{ R^{\delta} }.
\label{matching}
\end{equation}
A related (and the usual) way to derive this 
matching 
is to do a KK decomposition of
the canonically normalized
$( 4 + \delta ) D$ gauge field, i.e., with action $S = \int 
d^4 x d^{\delta} y \;
F^2_{\mu \nu} + 
\int d^4 x \; \bar{\psi} (x) \gamma ^{\mu} A_{\mu} (x, y=0) \psi(x) \;
g_{( 4 + \delta ) D} + ..$, where $y$ denotes the extra dimensions 
and we have 
assumed that the matter field $\psi$ is localized
at $y =0$:
the gauge boson in $( 4 + \delta ) D$ has 
mass dimension $1 + \delta /2$.
The zero-mode of the KK decomposition,
$A_{\mu}^{(0)} (x)$ with mass dimension $1$,
is the usual $4D$ gauge field and its
wavefunction has a 
normalization factor (from the volume of the extra dimensional space)
$
\sim 1/ \sqrt{R^{\delta}}$ 
so that the
coupling 
of zero-mode gauge boson
to matter field is $\bar{\psi} \gamma ^{\mu} A^{(0)}_{\mu} \psi \; 
g_{( 4 + \delta ) D} / \sqrt{R^{\delta}}$ and hence we get
the above result.
\footnote{
In a supersymmetric theory
with holomorphic normalization, the action is
$S = \int d^4 x d^{\delta} y \; 1/ g_{( 4 + \delta ) D}^2 F_{\mu \nu}^2
+ \int d^4 x \; \bar{ \psi } (x) A_{\mu} (x , y=0) \psi (x) + ..$, where
the gauge boson has mass dimension $1$ and $g_{( 4 + \delta ) D}^2$ has
mass dimension $-\delta$. Since the zero-mode is the Fourier component
of $( 4 + \delta ) D$ gauge field which is a constant
function of the extra dimensional coordinate, 
integration over the coordinate of the extra dimension
(to get $4D$ action)
gives simply a volume factor in the kinetic term for
the zero-mode, 
i.e.,
$S= \int d^4 x \left[
R^{\delta} / g_{( 4 + \delta ) D}^2 \left( F^{(0)} _{\mu \nu} 
\right)^2 + \bar{ \psi } \gamma ^{\mu} A_{\mu}^{(0)} \psi  + ..\right. $.
Hence we get above relation for the (holomorphic) gauge couplings.
} 
Thus, this 
argument justifies the above matching relation (Eq. (\ref{matching}))
at {\em tree}-level (i.e., {\em without} the energy dependence). 
We claim that this relation is valid at the {\em quantum}-level.

\section{Strong Coupling and Need for a Cut-off}
\label{cut-off}
We now review the relationship between the compactification
scale and the strong coupling scale in these two approaches. 

Consider $U(1)$ case where $e_{4D}$ has a logarithmic (mild)
energy dependence. We see that $e_{4D}^{\hbox{eff}}$ (Eq.
(\ref{eeff})) reaches
strong coupling, i.e., $e_{4D}^{\hbox{eff} \; 2} / \left(
16 \pi^2 \right) \sim O(1)$ (so that loop corrections become $\sim
O(1)$ or $\sim$ tree-level terms)
at an energy
$M$ such that $M ^{\delta} \sim R^{-\delta} \times \left(
16 \pi^2 \right)$ (assuming $e_{4D} \sim O(1)$ at $R^{-1}$).

From $( 4 + \delta ) D$ point of view, the gauge
theory is non-renormalizable (gauge coupling is dimensionful and so
there are power divergences in,
say, electron wavefunction
renormalization as shown above) and so we need a cut-off, say $M$.
The value of $e^2_{( 4 + \delta ) D}$ at strong coupling
(i.e., the maximum value for $e^2_{( 4 + \delta ) D}$) is
$\sim l_{4 + \delta} / M^{\delta} $,
where $l_D \equiv 2^D \pi ^{D/2} \Gamma (D/2)$,
such that $1/l_D$ is the loop expansion parameter 
(loop-factor, for short) in $D$
dimensions (for example, $1/ \left( 16 \pi ^2 \right)$ in $4D$)
\cite{clp}. Thus, the maximum value for
$e^2_{4D}$ (from Eq. (\ref{matching})) is
\footnote{In the $U(1)$ case, $e_{( 4 + \delta ) D}$ does not run 
and so its 
value at the compactification scale $\sim R^{-1}$ is the same as at
$M$.} 
$\sim 
l_{4 + \delta} / (R^{\delta} M
^{\delta})$ and so
if we require $e_{4D} \sim O(1)$, then we need $M 
^{\delta} \stackrel{<}{\sim} l_{4 + \delta} \; R^{-\delta} $, i.e.,
in agreement with above, we see that that we cannot 
(perturbatively) extrapolate the theory
to energies larger than $\sim \left( 1/ \hbox{loop-factor} \times
R^{-1} \right)$.

The non-abelian case is a 
bit subtle since $g_{4D}$ (or $g_{( 4 + \delta ) D}$)
also (in addition to $N_{KK} (E)$)
has a power-law dependence on $E$ due to running (at one-loop).  
In fact, with no matter fields in bulk, this effect {\em decreases}
$g_{4D}$ at higher energies 
(as shown below)
and thus competes with
$N_{KK}$ in determining how $g^{\hbox{eff}}_{4D}$ depends on energy
(see $4D$ side of Eq. (\ref{dict}) and below).

We now calculate the power-law running of $g_{4D}$.
In the
renormalizable approximation,
we have a finite number of KK 
states (up to
a cut-off $\Lambda$) and we can
treat the KK states as thresholds for
running of couplings, i.e., 
in the RGE, we decouple the KK states at their masses
$\sim n / R$. Thus, at one-loop,
we get the following 
RGE,
neglecting the effects of the zero-mode of the gauge field and 
matter fields:
\begin{equation}
\frac{ \partial g^{-2} _{4D} (E) }{ \partial \ln E}
\approx - \frac{ b _{KK} }{ 8 \pi ^2 } N_{KK} (E),
\label{rge}
\end{equation}
where $N_{KK} (E) \sim E^{\delta}
R^{\delta}$ is the number of KK states lighter than
$E$
({\em excluding} the zero-modes)
and $b_{KK} < 0$ is the
$\beta$-function coefficient for KK states at each (massive) level. 
To be precise, let us assume that the $\delta$ extra dimensions are
compactified on circles of equal radii $R$ so that the mass splitting 
between
KK states is $\approx 1/R$. We also assume $E \gg 1/R$
so that the sum over KK states can be approximated
by an integral. Then, 
$N_{KK} (E)$ is given by the volume of
$\delta$-dimensional sphere
of radius $ER$ (which is the maximum quantum number of the KK states),
but {\em not} counting the zero-mode, i.e.,
\begin{equation}
N_{KK} (E) \approx \hat{V}_{\delta} (ER)^{\delta} - 1,
\label{NKK}
\end{equation}
where 
$\hat{V}_{\delta} = 1 / \delta
\times
2 \pi ^{\delta /2} / \Gamma \left( \delta /2 \right)$
is the volume of a unit-sphere in $\delta$ dimensions. 

Integrating the RGE in Eq. (\ref{rge}) (using Eq. (\ref{NKK})) from 
$R^{-1}$ to energy scale $E$ (``bottom-up'' calculation
as in \cite{ddg}), we get
the gauge coupling at $E$ in terms of the gauge coupling at
$R^{-1}$:
\begin{eqnarray}
g^{-2}_{4D} (E) & \approx &
g^{-2}_{4D} ( R^{-1} ) - b_{KK} / 8 \pi^2 
\; \left[ \left(
( E R ) ^{\delta} - 1 \right) \hat{V}_{\delta}
/ \delta - \ln (ER) \right] 
\nonumber \\
 & \approx & g^{-2}_{4D} ( R^{-1} ) - b_{KK} / 8 \pi^2 \;
\left[ N_{KK} (E) / \delta + \left( 1 - \hat{V}_{\delta} \right) / \delta
- \ln (ER) \right].
\label{rgesol1}
\end{eqnarray}
We see that $g_{4D}$ {\em decreases}
(at one-loop)
with a power-law
as the 
energy is increased as expected since the (massive) gauge field KK states
make the theory {\em more} asymptotically free.
Thus, the effective gauge coupling is
\begin{equation}
\frac{ g^{\hbox{eff} \; 2}_{4D} (E) }{ 16 \pi^2 }
\approx \frac{ g^2_{4D} (R^{-1}) / 16 \pi^2  \times N_{KK} (E) }{
1 - b_{KK} / 8 \pi^2 \; \left[ N_{KK} (E) / \delta +
\left( 1 - \hat{V}_{\delta} \right) / \delta - \ln (ER) \right] \; 
g_{4D}^2
(R^{-1}) }.
\label{geff} 
\end{equation}
We can trace the $\ln (ER)$ factor in the equations 
above to the fact that the zero-mode of the gauge field
does {\em not} have the $\beta$-function coefficient
$b_{KK}$. As mentioned earlier,
we neglect the effect of the zero-mode and also of matter fields --
strictly speaking there
should be an additional term $\propto b_0 \ln (ER)$
in the above equations, where
$b_0$ is the 
$\beta$-function coefficient of zero-modes (zero-mode of gauge 
field $+$ matter fields), such that if $b_{KK} = b_0$, then these two
$\ln (ER)$ terms cancel each other.
As mentioned before,
the expression for $N_{KK} (E)$ in Eq.
(\ref{NKK}) and hence the solution to the RGE is really 
valid only for $ER \gg 1$, in 
which case the $(ER)^{\delta}$ factor dominates the 
$\ln (ER)$ factor in the above equations
and hence the latter can be neglected. Of course, 
the zero-modes also renormalize $g_{4D}$ below $R^{-1}$ as usual.

From Eq. (\ref{geff}) (and using 
(\ref{NKK})) we see that at an energy $M$ given by $M^{\delta}
\sim 8 \pi^2 / g^2_{4D} (R^{-1}) R^{-\delta}
\times - \delta / b_{KK}
$, the power-law term $\propto N_{KK}$ (from the running of $g_{4D}$)
starts to dominate in
the {\em denominator} in Eq. (\ref{geff}) (i.e., it becomes $O(1)$)
which implies that for $E\stackrel{>}{\sim} M$
the power-law running of $g_{4D}$ 
``cancels'' the power-law energy dependence of $N_{KK}$ in the {\em 
numerator}
in Eq. (\ref{geff}).
But, we see that at $E \sim M$,
$g_{4D}^{\hbox{eff} \; 2} / \left( 16 \pi^2 \right) \sim 
O \left( - \delta / (2 \; b_{KK}) \right)$
and also that $g^{\hbox{eff} \; 2}_{4D}
/ \left( 16 \pi^2 \right)$ reaches a {\em constant} value,
$- \delta / (2 \; b_{KK})$,
as $E \rightarrow \infty$. Thus, by the energy scale 
$M$ at which
$g^{\hbox{eff} \; 2}_{4D}$ starts ``leveling'' off, 
we see that it has already reached
strong coupling.
In other words, 
even though $g^{\hbox{eff} \; 2}_{4D}$
does {\em not} grow ``indefinitely'' with energy
(unlike in the $U(1)$ case: Eq. (\ref{eeff})), 
the theory becomes non-perturbative above $
M \sim R^{-1} \times \left( 1/ \hbox{loop-factor}
\right)^{1 / \delta}$ as in the $U(1)$ case.
\footnote{At strong coupling, one might worry about {\em (power-law) 
higher}-loop corrections
to the gauge coupling and thus whether 
the one-loop RGE for the gauge coupling
suffices. 
However, if the massive gauge KK states
form $N=2$ SUSY vector multiplets (as in \cite{ddg}), then there are {\em no}
corrections 
(power-law or logarithmic) beyond one-loop involving {\em only 
massive} gauge KK
states. Of course, 
at two-loop, the gauge coupling depends on
the wavefunction renormalization of matter fields 
which are on $3$-branes (and also on wavefunction renormalization of
{\em zero-}mode 
gauge fields which do {\em not} form $N=2$ SUSY vector multiplets)
which, in turn, evolves like a power of energy due to the KK states
of gauge fields (as discussed earlier). However, this {\em two-loop} 
power-law
running of the
gauge coupling is clearly suppressed by the usual
loop-factor $\sim g_{4D}^2 / \left( 16 \pi^2 \right)$ 
compared to the {\em one-loop} power-law running. 
Thus, the one-loop RGE for the gauge coupling
suffices (for our purpose), even at strong coupling. Of course,
wavefunction renormalization of matter fields
gets a contribution from the massive KK states to {\em all} loop
orders and hence, at strong coupling,  
the one-loop RGE will {\em not} suffice for evaluating it,
i.e., the $n$-loop term $\sim 
\left[ N_{KK} \; g_{4D}^2 / \left( 16 \pi^2 \right) \right] ^ n$
is $O(1)$ at strong coupling.
}

To complete this discussion, we have to look at the relation between the
strong coupling scale
and the compactification scale in the non-abelian case {\em from
the} 
$( 4 + \delta ) D$ {\em point of view}. The argument is similar to 
that in the $U(1)$
case, except that   
we have to run $g_{( 4 + \delta ) D} (E)$ from the cut-off ($M$) to $R^{-1}$
\footnote{The reader might still be 
uncomfortable with the terminology
``running'' of gauge coupling in $( 4 + \delta ) D$ -- in that case,
a better term is
``finite energy-dependent corrections'' to the gauge coupling.} 
and then match
to the (effective) $4D$ theory 
to give $g_{4D} ( R^{-1} )$. Although, as before, 
we refrain from
doing this $( 4 + \delta ) D$ calculation, it should agree with the above
calculation in the 
$4D$ theory with KK states.      

\section{Gaugino Masses in
Kaluza-Klein/Radion Mediated SUSY Breaking}
Next, we use the discussion in the previous sections
to resolve the discrepancy in the results for
gaugino masses in KKMSB \cite{ky} and RMSB \cite{cl}.

Suppose SUSY is broken by the
radion -- to repeat this
is the field whose vev
determines the size $R$ of the extra dimension(s). 
In general,
to compute (zero-mode)
gaugino masses, we have to determine how the low energy
($4D$) gauge couplings
(in other words, the wavefunction renormalization of gauge fields)
depend on the SUSY breaking modulus
\cite{gr}. 
Thus, in this case,
we need to integrate the RGE
(Eq. (\ref{rge}))
starting from cut-off $\Lambda$ (``top-down'' approach; see,
for example, \cite{tv, ross})
and compute the dependence of
$g_{4D} (R^{-1})$ on $R$ -- this just amounts to setting $E \approx 
\Lambda$
in Eq. (\ref{rgesol1}) and rewriting it to get
\begin{eqnarray}
g_{4D}^{-2} ( \mu \sim R^{-1} ) & \approx & g^{-2}_{4D} (\Lambda)
- \frac{b_{KK}}{8 \pi^2} \left[ \frac{  \hat{V_{\delta}} }
{\delta}
\left( 1 - ( \Lambda \; R )^{\delta} 
\right) + \ln ( \Lambda R ) \right] 
\label{rgesol2}
\end{eqnarray}
with 
\begin{equation}
g^{-2}_{4D} (\Lambda) \approx
g^{-2}_{( 4 + \delta ) D} (\Lambda) \left( 2 \pi R
\right) ^{\delta}
\label{bc}
\end{equation}
obtained from the matching condition, Eq. (\ref{matching})
(we have added factors of $2 \pi$ in the extra dimensional volume). 
Here, we have neglected the effects of the zero-mode of the gauge field
and also of matter fields (which are asssumed to be on $3$-branes)
on the running. 
%\footnote{
The above result is the same as
Eq. (32) in \cite{ross}.
As mentioned earlier, the
solution in Eq.
(\ref{rgesol2}) is strictly speaking valid only for
$\Lambda R \gg1$; in this case, the $\ln ( \Lambda R )$ term
can be neglected compared to the $\left( \Lambda R \right )^{\delta}$ term.
%}
Although, the expression for $g_{4D} \left(
\mu \sim R^{-1} \right)$ in Eq. (\ref{rgesol2})
has been obtained using the $4D$ approach, 
it is clear from the discussion in the previous sections that the
$5D$ approach will give the same expression (with
$g^{-2}_{4D} ( \Lambda )$ given by 
Eq. (\ref{bc})).

The gaugino mass is given by
\begin{equation}
M_{\tilde{g}} ( \mu \sim R^{-1} ) \approx F_T \; g^2 _{4D} (
\mu \sim R^{-1} ) \frac{ \partial g^{-2}_{4D} (
\mu \sim T ) }{ \partial T } | _{ T \sim R^{-1} },
\label{mass}
\end{equation}
where $T$ is the canonically normalized
radion chiral superfield, i.e.,
$\langle T \rangle \sim R^{-1} + F_T \theta ^2$
and $g^{-2}_{4D} (\mu
\sim T)$ is the SUSY generalization of Eqs. (\ref{rgesol2}) and (\ref{bc}).
\footnote{
A brief comment on the effect of zero-modes
(zero-mode of gauge field
and also matter fields) on the gaugino mass is in order here.
As mentioned earlier,
the running due to zero-modes 
will result in an additional term
$\propto b_0
\ln ( \Lambda R )$ in $g^{-2}_{4D}
\left( \mu \sim R^{-1} \right)$ and thus seems to give an additional
(weak) dependence on $T$ when computing $\partial g^{-2}_{4D} (
\mu \sim T ) / \partial T$ in the above equation. However,
it is clear that this dependence
``cancels'' when we run from $R^{-1}$ to the weak scale,
i.e., the running contribution due to zero-modes (unlike the KK modes)
to $g^{-2}_{4D} ( \mu \sim weak \; scale)$ obviously does {\em not} 
depend on $R$. 
Of course, 
the zero-modes 
do affect
the gaugino mass since, at one-loop, 
$M_{\tilde{g}} / g^2$ is RG-invariant and zero-modes 
renormalize (as in $4D$) $g_{4D}$
from $\Lambda$ to the weak scale
-- this effect on
gaugino mass in running 
from
$\Lambda$ to $R^{-1}$ 
will appear in the $g^2_{4D} \left( \mu \sim R^{-1} 
\right)$ term in Eq. (\ref{mass}) and the RG effect from
$R^{-1}$ to the weak scale 
(not shown here) is the usual ($4D$) running of gaugino mass.}
To be precise, $g^{-2} _{4D} (\mu \sim T)$ used to compute the
derivative in
Eq. (\ref{mass}) is the holomorphic
gauge coupling 
(including the topological vacuum angle, i.e., the
$\theta$-term), whereas
$g^{-2}
_{4D} \left( \mu \sim R^{-1} \right)$ in Eq. (\ref{rgesol2}) is 
(closer to)
the physical or canonical gauge coupling. 
In general, the canonical gauge coupling differs
from the holomorphic gauge coupling due to anomalous Jacobians under the
rescaling of gauge fields 
in going from holomorphic to canonical normalization 
\cite{nima}.
Suppose the {\em massive} gauge KK states 
(at each level) form $N = 2$ SUSY vector
multiplets (as in \cite{ddg}): in the $N=1$ SUSY language, these consist 
of a vector multiplet and a chiral multiplet in the adjoint
representation. The anomalous Jacobians
from these two $N=1$ SUSY multiplets {\em cancel} each other \cite{nima}.
\footnote{By $N=2$ supersymmetry, at each massive level, the rescalings,
i.e., the wavefunction renormalization, for the 
chiral and vector multiplets 
must be the same (up to the loop effect of {\em zero}-mode gauge fields
which do {\em not} form $N=2$ SUSY multiplet).} 
%\footnote{
This cancelation is related to the fact that there
are no corrections to the canonical gauge coupling
{\em beyond}
one-loop involving {\em only massive} gauge KK
states (i.e., $N=2$ SUSY vector multiplets) 
\cite{nima}.
%}
The {\em zero}-modes of the gauge fields
form a $N=1$ SUSY vector multiplet (as usual)
which {\em does} give an anomalous Jacobian under the rescaling, and hence
the following relation (for an $SU(N)$ gauge group): 
$ 1/ g_{4D \;c}^2 (E) = \hbox{Re} \left( 1 / g_{4D \; h}^2 (E)
\right) - 2 N / 
\left( 8 \pi^2 \right) \ln g_{4D \; c} (E)$ \cite{shifman,
nima}, where the subscript $c$ ($h$)
denotes the canonical (holomorphic) gauge coupling.
Thus, the {\em RG scaling or running} of the (real part of) gauge coupling 
in these two 
normalizations differs only at 
{\em two} (and higher)-loop level. 
So, the {\em one}-loop result for the canonical gauge coupling, $g^{-2}
_{4D} \left( \mu \sim R^{-1} \right)$, 
can be generalized to the 
holomorphic gauge coupling,
i.e., to $g^{-2} _{4D} (\mu \sim T)$, 
by the simple substitution $R^{-1} \rightarrow T$ as required by holomorphy.
Of course, the canonical
gauge coupling at the cut-off, $g^{-2}_{4D \; c} (\Lambda)$, differs from
the holomorphic gauge coupling,
$g^{-2}_{4D \; h} (\Lambda) \sim g^{-2}_{(4 + \delta) D \; h} (\Lambda)
\; T^{\delta}$, by the ``one-loop'' 
term from the rescaling anomaly, $\sim
- 2 N /
\left( 8 \pi^2 \right) \ln \left( R ^{- \delta /2} \right)$
(using $g_{4D \; c} \propto R^{- \delta /2}$);
this results in an additional (mild) dependence of the canonical gauge
coupling on $R$. \footnote{There are loop corrections to the kinetic
terms of 
matter fields which are on $3$-branes
(wavefunction renormalization $Z$)
and thus there is also an anomalous Jacobian under 
rescaling of matter fields in going to canonical kinetic terms. As in
the case of gauge fields,
this rescaling modifies the RGE's for gauge couplings only at {\em two}-loop
level
and hence does not modify the {\em one}-loop analysis above. 
Also, since the matter fields are on $3$-branes, the kinetic
terms of matter fields 
do not depend on $R$ at {\em tree-level} 
and hence the
rescaling anomaly term $\sim 1/ \left( 8 \pi^2 \right) \ln Z$ does
{\em not} depend on  
$R$ at the one-loop level (unlike in the case of
rescaling of gauge fields: see above).} 

The low energy ($4D$) gauge coupling depends on $R$ due to two
effects. One dependence of $g_{4D} \left( \mu \sim R^{-1}
\right)$ on $R$ in Eq. (\ref{rgesol2})
is from the (one-loop) power-law
running as discussed by
Kobayashi and Yoshioka (KY) \cite{ky}; this is the effect of quantum 
power-law scaling.
This dependence gives 
\begin{eqnarray}
\frac{\partial
g^{-2}_{4D} \left( \mu \sim T \right)}{\partial T} | _{KY}
& \approx & - \frac{ b_{KK} }{ 8 \pi^2 } R \left[
\hat{V_{\delta}}
( \Lambda R )^{\delta} - 1 \right]
\nonumber \\
 & \approx & - \delta \; R \times
\; \hbox{running contribution to} \; g_{4D}^{-2} \left(
\mu \sim R^{-1} \right) \nonumber \\
 & & \hbox{
(up to a small log-factor, see Eq. (\ref{rgesol2}))}.
\label{derky}
\end{eqnarray}
Here, increasing $R$ makes $g_{4D}
(\mu \sim R^{-1})$
larger due to larger number of KK states contributing to running
(with $b_{KK} < 0$)
so that the above
derivative
has positive sign.

%Kobayashi and Yoshioka (KY) \cite{ky}, using the ``$4D$'' approach, 
%assume that $g^{-2}_{4D} (\Lambda)$ is a ``fundamental'' parameter and 
%hence has no functional dependence on $R$ (i.e., $T$) so that 
%the only dependence of $g_{4D} \left( \mu \sim R^{-1} \right)$ on $R$ 
%in Eq. (\ref{rgesol2}) is from the power-law running; this is the 
%effect of quantum scaling. 
%Then, we get  
%\begin{eqnarray} \frac{\partial g^{-2}_{4D} \left( \mu \sim T \right)}
%{\partial T} | _{KY} & \approx & - \frac{ b_{KK} }{ 8 \pi^2 } R \left[
%\hat{V_{\delta}}
%( \Lambda R )^{\delta} - 1 \right]
%\nonumber \\
% & \approx & - \delta \; R \times
%\; \hbox{running contribution to} \; g_{4D}^{-2} \left(
%\mu \sim R^{-1} \right) \nonumber \\
% & & \hbox{
%(up to a small log-factor, see Eq. (\ref{rgesol2}))}.
%\label{derky}
%\end{eqnarray}
%%
%In the KY case (with fixed $g_{4D} (\Lambda)$), 
%increasing $R$ makes $g_{4D}
%(\mu \sim R^{-1})$
%larger due to larger number of KK states contributing to running
%(with $b_{KK} < 0$) 
%so that the above
%derivative
%has positive sign. 
%This gives \cite{ky}
%%
%\begin{eqnarray}
%M_{\tilde{g}} (\mu \sim R^{-1}) |_{KY} & \approx & F_T R \times -
%\frac{ b_{KK} \; g^2_{4D} (\mu \sim R^{-1}) }{ 8 \pi ^2 }
%\left[ \hat{V}_{\delta} (  \Lambda R ) ^{\delta} - 1 \right]
%\nonumber \\
% & \approx & F_T R \times -
%\frac{ b_{KK} \; g^2_{4D} (\mu \sim R^{-1}) }{ 8 \pi ^2 } \;
%N_{KK} (\Lambda),
%\label{kymass}
%\end{eqnarray}
%
%where $N_{KK} (\Lambda)$ is the total number of KK states (up to
%the cut-off).

The other dependence of 
$g_{4D} \left( \mu \sim R^{-1}
\right)$ on $R$ is from the value of $g_{4D}$ at the cut-off $\Lambda$.
The $4D$ theory with KK states is derived from the (``fundamental'')
$( 4 + \delta )$ theory. Hence, $g^{-2}_{( 4 + \delta ) D} (\Lambda)$
is a fundamental parameter 
so that (SUSY generalization of) $g^{-2}_{4D} (\Lambda)$ depends on $T$
(see Eq. (\ref{bc})) as discussed by Chacko and Luty (CL) \cite{cl}.
CL use the $5D$ approach in which this dependence is obvious,
\footnote{$g^{-2}_{( 4 + \delta ) D}
(\Lambda)$ is determined by, say, the
dilaton field $\phi$
in $( 4 + \delta ) D$ (in the context of
string theory). Thus, $g^{-2}_{4D} (\Lambda)$ depends on a
combination of the fields
$\phi$ and radion ($R$). We can {\em define} (the real part of)
the dilaton chiral superfield $S$ {\em in} $4D$ to be this combination of
$\phi$ and $R$ \cite{witten}
so that we get the
tree-level expression $g^{-2}_{4D} \sim \hbox{Re}S$
which is commonly used in
the literature.} but it is clear that the same effect appears in the $4D$
approach as well.
This
effect
corresponds to
classical power-law scaling in the sense that
this dependence of $g^{-2}_{4D} (\Lambda)$ (and hence of
$g^{-2}_{4D} \left( \mu
\sim R^{-1} \right)$) on $R$ is present
in $U(1)$ case also, i.e.,
it is a tree-level and {\em not} the running
effect.
% 
%Whereas Chacko and Luty (CL) \cite{cl}, 
%using the ``$5D$'' approach,
%assume that $g^{-2}_{( 4 + \delta ) D} (\Lambda)$ (instead of $g^{-2}_{4
%D} (\Lambda)$)
%is fundamental 
%\footnote{$g^{-2}_{( 4 + \delta ) D}
%(\Lambda)$ is determined by, say, the 
%``dilaton'' field $\phi$
%in $( 4 + \delta ) D$ (in the context of
%string theory). Thus, $g^{-2}_{4D} (\Lambda)$ depends on a
%combination of the fields
%$\phi$ and radion ($R$). We can {\em define} (the real part of)
%the dilaton chiral superfield $S$ {\em in} $4D$ to be this combination of
%$\phi$ and $R$ \cite{witten}
%so that we get the 
%tree-level expression $g^{-2}_{4D} \sim \hbox{Re}S$ 
%which is commonly used in
%the literature.} and hence 
%(SUSY generalization of) $g^{-2}_{4D} (\Lambda)$ depends on $T$
%(see Eq. (\ref{bc}));
%this corresponds to
%classical scaling in the sense that
%this dependence of $g^{-2}_{4D} (\Lambda)$ (and hence of
%$g^{-2}_{4D} \left( \mu
%\sim R^{-1} \right)$) on $R$ is present
%in $U(1)$ case also, i.e.,
%it is a tree-level and {\em not} the running
%effect. In this case, we get an {\em additional} contribution
%
Thus, we get an {\em additional} contribution
\begin{eqnarray}
\frac{ \partial g^{-2} _{4D} \left( \mu \sim T \right) }{\partial T}
|{b.c.} & \approx - \delta \; R \; 
g^{-2}_{( 4 + \delta ) D} (\Lambda) \left( 2 \pi R
\right) ^{\delta} & \approx
- \delta \; R \times g^{-2}_{4D} (\Lambda),
\label{derbc}
\end{eqnarray}
where b.c. stands for ``boundary condition''.
Since larger $R$ makes $g^{2}_{4D}
(\Lambda)$ (and hence $g^{2}_{4D}
(\mu \sim R^{-1})$) smaller for fixed $g^{2}_{( 4 + \delta ) D}
(\Lambda)$, this contribution to the above
derivative is of negative sign.

The contribution to the derivative in Eq. (\ref{derbc})
is always larger (in magnitude) than the one in Eq.
(\ref{derky}) since
to keep $g^{-2}_{4D} (\mu \sim R^{-1})$ (Eq. (\ref{rgesol2}))
positive, i.e., to prevent
$g_{4D} (\mu \sim R^{-1})$ from ``blowing'' up, the running 
(i.e., loop) contribution
to $g^{-2}_{4D} (\mu \sim R^{-1})$ (which enters in Eq.
(\ref{derky})) has to be smaller in magnitude
than the value at $\Lambda$ (which enters in Eq. (\ref{derbc})).
It is clear from Eqs. (\ref{rgesol2}), (\ref{derky}) and (\ref{derbc}) that
if the suppression due to the loop-factor
in the running contribution is compensated by either 1)
$g_{4D} \gg 1$ 
at the cut-off (i.e., $g^{-2}_{4D} ( \Lambda )$ is {\em small})
or
by 2) a large number of KK states (i.e., $\Lambda R \gg 1$; in this
case, $g_{4D} (\Lambda)$ can be $O(1)$),
then 
the two contributions
to the derivative (and hence
to gaugino mass)
can be {\em comparable (in magnitude)}.
The first case is ruled out since $g_{4D} (\mu \sim R^{-1})$ (and hence
$g_{4D} (\mu \sim
\hbox{weak scale})$) will also be much larger than $1$,
whereas we know that the measured SM gauge couplings are all at most $O(1)$.
In either of these two cases, we see that
$g^{-2} _{( 4 + \delta ) D} (\Lambda) \sim O 
\left( - b_{KK} / ( 8 \pi^2 ) \Lambda ^
{\delta} \right) \ll
\Lambda ^ {\delta}$, i.e., the $( 4 + \delta ) D$ theory 
is strongly coupled at
the cut-off $\Lambda$ -- this is expected since from the $( 4 + \delta ) D$
theory point of view, the only way that the running effect can be as
important as the tree-level effect is for the theory
to be strongly coupled. 
%\footnote{
However, the $4D$ theory
can still be weakly coupled (at the cut-off)
if $\Lambda R \gg 1$ as discussed above (case 2) and as seen from
Eq. (\ref{bc}).
%}

Reference \cite{ky}
uses the $4D$ approach, but the 
tree-level contribution to the gaugino mass
(i.e, the effect of classical power-law scaling),
Eq. (\ref{derbc}), is {\em not} included -- in other words, it is 
assumed that $g^{-2}_{4D} (\Lambda)$ is a ``fundamental'' parameter
and hence has no functional dependence on $R$.
Then,
the running 
(or quantum power-law scaling)
contribution, Eq. (\ref{derky}), (by itself) gives \cite{ky}
\begin{eqnarray}
M_{\tilde{g}} (\mu \sim R^{-1}) |_{KY} & \approx & F_T R \times -
\frac{ b_{KK} \; g^2_{4D} (\mu \sim R^{-1}) }{ 8 \pi ^2 }
\left[ \hat{V}_{\delta} (  \Lambda R ) ^{\delta} - 1 \right]
\nonumber \\
 & \approx & F_T R \times -
\frac{ b_{KK} \; g^2_{4D} (\mu \sim R^{-1}) }{ 8 \pi ^2 } \;
N_{KK} (\Lambda),
\label{kymass}
\end{eqnarray}
where $N_{KK} (\Lambda)$ is the total number of KK states (up to
the cut-off).

Whereas, {\em adding} the contributions in
Eqs. (\ref{derky}) and (\ref{derbc})
and using Eq. (\ref{rgesol2}), we get
\begin{eqnarray}
M_{\tilde{g}} (\mu \sim R^{-1})
& \approx & - \delta F_{T} R \left[ 1
+  \frac{b_{KK}}{8 \pi^2} g_{4D}^2 \left( \mu \sim R^{-1} \right)
\left( ( \hat{V}_{\delta} -1) / \delta + \ln (\Lambda R) \right) \right].
\label{clmass}
\end{eqnarray}
It is clear from the above discussion that the 
KY result in Eq. (\ref{kymass}) is smaller than
(or at most comparable to,
as argued above) 
the  
above result and also of opposite
sign. 
In any case, it is clear that according to KY,
the gaugino mass is a {\em loop-level} effect
(which, however, can be {\em enhanced} by a large number of 
KK states), whereas 
CL argue that it is (also) a {\em tree-level} effect.

In the $U(1)$ case
with matter fields on $3$-brane,
the discrepancy between the two results is even more obvious.
According
to KY, $e_{4D}$ does {\em not} depend on $T$
(since there is no running due to KK states: $b_{KK} = 0$) so that
there is {\em no} photino mass (at one-loop),
whereas CL show that there {\em is} photino mass:
in the $4D$ approach, this is
due to the dependence of $e_{4D}^{-2} (\Lambda)$
on $T$. 
%(i.e., due to the boundary condition, Eq. (\ref{bc})).
To repeat, the 
resolution of this discrepancy (which also
applies to the non-abelian case) is that the boundary (or
tree-level) conditions
in the two cases are different -- KY assume that
$g^{-2}_{4D} (\Lambda)$ is fundamental (no
functional dependence on $R$),
whereas CL assume $g^{-2}_{( 4 + \delta ) D} (\Lambda)$
is fundamental and hence
$g^{-2}_{4D} (\Lambda)$ {\em does} depend on $R$ (Eq. (\ref{bc})).

Actually, CL did not
{\em explicitly} 
include (one-loop) power-law running of $g_{( 4 + \delta ) D}$ (or
equivalently that of $g_{4D}$), i.e., the effect of quantum 
power-law scaling
%they assumed that $g^{-2}_{( 4 + \delta ) D} \left( \mu \sim R^{-1} \right)$
%is {\em in}dependent of $R$ 
-- this corresponds to assuming
$b_{KK} = 0$ in our notation.
So they get
$M_{\tilde{g}} (\mu \sim R^{-1}) |_{CL}
\approx - \delta F_{T} R$
\cite{cl}
(i.e., Eq. (\ref{clmass}) with $b_{KK} = 0$). This is of course valid for 
$U(1)$ case. In the non-abelian case, 
as argued earlier, it is possible that the running (loop)
contribution to gaugino mass (Eq. (\ref{derky})) 
can be of the {\em same order} as the tree-level
contribution (Eq. (\ref{derbc}))
if the $( 4 + \delta ) D$ theory is strongly coupled at
$\Lambda$. However, in the non-abelian case, 
we can see
that the 
expression for the gaugino mass is approximately
the same (as in the $U(1)$ case) since 
(with $b_{KK} \neq 0$) the second term 
in the bracket in
Eq. (\ref{clmass}) (including the $\ln ( \Lambda R )$ piece) 
is small. 
%\footnote{
This is true even if $\left( \Lambda R \right)^{\delta}
\sim 1/$loop-factor $\gg 1$,
which, as discussed in section \ref{cut-off}, corresponds to the
maximum value of $\Lambda$. Here, we assume
$g_{4D} \left( \mu
\sim R^{-1} \right) \stackrel{<}{\sim} O(1)$ as in the MSSM. 
Explicitly, with $b_{KK} \neq 0$, it is obvious that in
Eq. (\ref{mass}) there are
extra terms (as compared to $U(1)$ case) 
in {\em both} $g_{4D}^{-2} (\mu \sim R^{-1})$
(see Eq. (\ref{rgesol2})) and $\partial g^{-2}_{4D} 
( \mu \sim T ) / \partial T$ (Eq.
(\ref{derky})) which (almost) {\em cancel} each other and
give the same expression for the gaugino mass
as in the $U(1)$ case.
\footnote{
In other words, for $\Lambda R \gg 1$, we get from Eqs.
(\ref{rgesol2}) and (\ref{bc})
$g^{-2}_{4D} \left( \mu \sim R^{-1} \right)
\approx \left( 2 \pi R \right) ^{\delta} 
\left[ g^{-2}_{( 4 + \delta ) D} (\Lambda) + \hat{V}_{\delta}
/ ( 2 \pi ) ^{\delta} \;
b_{KK} / \left(
8 \pi^2 \right) \; \Lambda ^{\delta} / \delta \right]$
so that $g^2_{4D} \; \partial
g^{-2}_{4D} 
/ \partial T \approx - \delta R$ in Eq. (\ref{mass}) and hence
$M_{\tilde{g}} (\mu \sim R^{-1})
\approx - \delta F_T R $ (which is the expression in \cite{cl}), 
{\em irrespective} of the
values of 
$b_{KK}$ or $g_{( 4 + \delta ) D} (\Lambda)$.
In fact, assuming $g_{4D} (\Lambda) \sim O(1)$
(as in the MSSM), this corresponds to 
case 2 mentioned after Eq. (\ref{derbc}) --
the tree- and loop-level terms in the
above expression
for $g^{-2}_{4D} \left( \mu \sim R^{-1} \right)$ 
are {\em comparable}.  
The other possibility is that $\Lambda R \sim O(1)$ 
so that $R^{\delta}$ {\em cannot} be
``factored out'', unlike in the case $\Lambda R \gg 1$
(of course,
in this case, the result in Eq. (\ref{rgesol2}) may not
be valid as mentioned before). In this case, if the loop correction to the
gauge coupling is comparable to the value at the cut-off, then
we get $g^2_{4D} (\Lambda) \;
\left( \hbox{and also} \; g^2_{4D} \left
(\mu \sim R^{-1} \right) \right) \gg 1$ (corresponding to case 1 mentioned
before). This case {\em cannot} correspond to the MSSM, 
where all gauge couplings are $O(1)$. Therefore, if
$\Lambda R \sim O(1)$ in the MSSM, then the loop correction
to the
gauge coupling (and hence the running contribution to gaugino mass) 
has to be {\em small} so that we again recover the CL result.
}

%In the $U(1)$ case
%with matter fields on $3$-brane, there is no photino mass (at one-loop)
%according
%to KY since in the absence of running due to KK states
%$e_{4D}$ does {\em not} depend on $T$,
%whereas CL claim that there is photino mass
%since $e_{4D}^{-2} (\Lambda)$
%{\em does} depend on $T$. Thus, the discrepancy is
%even more
%obvious here and (from the above
%discussion) the resolution (which also
%applies to the non-abelian case) is that the boundary (or
%tree-level) conditions
%in the two cases are ``different'' -- to repeat, KY assume that
%$g^{-2}_{4D} (\Lambda)$ is fundamental (no
%functional dependence on $R$),
%whereas CL assume $g^{-2}_{( 4 + \delta ) D} (\Lambda)$
%is fundamental and hence
%$g^{-2}_{4D} (\Lambda)$ {\em does} depend on $R$ (Eq. (\ref{bc})).

\section{Summary}
In summary, we have reviewed the equivalence of the
``$4D$ with KK states'' and ``$5D$ with matching'' approaches
to study theories with gauge fields
in extra dimensions.
We reiterated that there are two different power-law scalings for
effective
gauge couplings which we referred to as classical and quantum. 

In supersymmetric theories with SUSY breaking in the radion,
the $4D$ approach {\em appears} to give gaugino mass at loop-level \cite{ky},
whereas the
$5D$ approach gives gaugino mass at tree-level \cite{cl}. We  
clarified that 
this 
discrepancy is due to 
the fact that, even in the $4D$ approach, there {\em is} a tree-level
contribution to gaugino mass due to the boundary condition, Eq. (\ref{bc}),
where $g^{-2}_{( 4 + \delta ) D } (\Lambda)$ is a fundamental parameter
and hence $g^{-2}_{4D} (\Lambda)$ depends on $R$. This contribution
was {\em not} included in \cite{ky} and it  
corresponds to the tree-level effect in \cite{cl}.
We also showed that 
even if the loop contributions to gauge coupling (and to gaugino mass) 
analyzed in \cite{ky} (and which are
{\em not} included in the {\em tree-level} analysis of \cite{cl}) 
are enhanced by large number of KK states
in the $4D$ approach (or equivalently,
due to strong coupling in $5D$ approach), the ``loop correction''
to the (tree-level) {\em expression} for gaugino mass given in
\cite{cl} is {\em small}. 

{\bf Acknowledgments}
This work is supported by DOE Grant DE-FG03-96ER40969.
The author thanks Nima Arkani-Hamed, 
Zackaria Chacko, Markus Luty and Martin Schmaltz for discussions,
Ann Nelson for suggesting the example in section
\ref{classical}
and 
the Aspen Center for Physics
for hospitality during the beginning of this work.

\end{document}